# Muochrony: Exploring Time and Frequency Applications of Cosmic Muons

G. Cerretto[1,6,*], E. Cantoni[1,6], M. Sellone[1,6], C.E. Calosso[1], I. Gnesi[2,3,4,6], H.K.M. Tanaka[5,6]
[1]Istituto Nazionale di Ricerca Metrologica (INRIM), Turin, Italy
[2]The University of Torino, Turin, Italy
[3]INFN LNF – Gruppo di Cosenza, Frascati (Rome), Italy
[4]CERN, Geneva, Switzerland
[5]The University of Tokyo, Tokyo, Japan
[6]International Virtual Muography Institute (VMI)
[*]Email: g.cerretto@inrim.it

*Summary*—This study outlines the progress of a collaborative effort between INRIM and MUOGRAPHIX-The University of Tokyo, focusing on using muons from cosmic-ray-induced Extensive Air Showers (EAS) to synchronize atomic clocks and disseminate atomic time references. The approach, known as the Cosmic Time Synchronizer (CTS), proposed by the University of Tokyo, serves as the foundation for a new field of study called Muochrony. The paper details the CTS technology, underlying principles, and the prototype system installed at the INRIM RadioNavigation Laboratory. Additionally, it reports on the initial metrological evaluation and the first experiments conducted to synchronize diverse atomic clock types and disseminate the UTC(IT) timescale using cosmic muons. CTS has the potential to synchronize and disseminate time references in critical applications securely and could also complement GNSS in areas not covered by RF signals.

*Keywords — Muography; timing; synchronization; dissemination; UTC*

I. INTRODUCTION

Modern technologies and automated systems are becoming increasingly complex, demanding higher levels of synchronization to ensure performance, security, and reliability. In recent years, stability in the microsecond and nanosecond ranges has become standard in various domains, including technology, infrastructure, finance, and daily life, driven in part by the widespread adoption of GPS/GNSS and the rollout of 5G networks. While atomic clocks and GNSS largely meet these stringent requirements, they also have notable limitations in terms of cost, reliability, and vulnerability to interference. High-performance atomic clocks are cost-prohibitive for many applications, and GNSS-based timing systems, though relatively affordable and accurate, are susceptible to jamming and spoofing, whether intentional or accidental, due to their reliance on RF signals. Additionally, these methods face significant challenges in environments such as indoor, underground, or underwater locations where RF signals are impractical.

Recognizing these limitations, international efforts in recent years have sought alternative solutions, exploring innovative methods for precise timing. One such approach is the Cosmic Time Synchronizer (CTS) [1], proposed by the University of Tokyo, which leverages muons from EAS [2] instead of RF signals from GNSS satellites, offering a robust and versatile alternative for timing applications, depending on the requirements of the final users.

II. PRINCIPLE OF FUNCTIONING

When cosmic rays enter Earth's atmosphere, they collide with air nuclei, triggering hadronic interactions that result in the generation of secondary particles, known as EAS. These showers are composed primarily of hadronic, electromagnetic, and muonic components, along with phenomena such as fluorescence, Cherenkov radiation, and neutrino production from particle decay. Each EAS, covering an average area of several square kilometers, delivers approximately $10^6$ secondary particles to the ground, with around 2% being relativistic muons, among the most penetrative of these particles, capable of traveling through air and solid material.

Due to their ability to penetrate dense objects, muons are widely used in imaging large structures like volcanoes and pyramids, a practice known as Muographic Imagery. Within an EAS, muons can be treated as nearly simultaneous events for detectors, with time differences in the order of tens of nanoseconds. Furthermore, the arrival time differences between detectors exhibit high temporal stability, influenced primarily by white Gaussian noise. These properties make muons highly suitable for precise timing applications, giving rise to the field of Muochrony, which parallels the established domain of Muographic Imagery and the emerging discipline of Muometry, focused on muon-based positioning and navigation.

Together, Muographic Imagery, Muometry (for positioning and navigation), and Muochrony (for timing) form the core branches of the broader scientific field of Muography [3], exploring unique and innovative applications of cosmic-ray-induced muons.

III. THE TECHNIQUE

A CTS network is composed of at least two sensors: a Master sensor linked to a time reference, such as an independent atomic clock, a GPS Disciplined Oscillator (GPS

DO), or a UTC(k) timescale, and one or more Slave sensors equipped with local clocks to be synchronized (e.g., Active Hydrogen Maser – AHM, Rubidium clocks or OCXOs). Each sensor records local measurements, assigning timestamps to detected muons based on its respective time reference. The Master sensor timestamps muons relative to the system's reference clock, while the Slave sensors use their local clocks. By analyzing and comparing timestamps from the Master and Slave sensors, coincidences in EAS muon detections are identified, allowing the time difference between the clocks to be calculated and synchronization achieved whenever a match is found.

Since muons are unaffected by electromagnetic interference and no critical timing data is exchanged between the Master and Slave sensors, CTS is inherently secure against jamming and spoofing attempts, whether accidental or intentional. A small-scale CTS network was implemented and tested at the University of Tokyo in 2022. Initial results, while preliminary, demonstrated promising synchronization stability of approximately 30 ns (1σ) over a 60 m range in an indoor environment with reinforced concrete barriers of 30 cm, simulating conditions where GNSS systems are impractical [4].

## IV. ASSESSMENT AT INRIM PREMISES

Building on the promising results from the University of Tokyo Laboratories, a collaboration was formalized between INRIM and VMI/MUOGRAPHIX- The University of Tokyo, leading to an actual experimental activity: the metrological evaluation of the CTS system at INRIM to assess its potential for time metrology applications.

As part of this effort, a CTS setup consisting of four 30 cm x 30 cm scintillators and associated measurement equipment was sent to INRIM and installed at its RadioNavigation Laboratory. This initial setup, currently undergoing testing, features CTS detectors positioned within a 5-meter radius. The first phase of testing involved connecting the Master and three Slave sensors to the UTC(IT) signal generated by INRIM's Time Laboratory to evaluate the CTS system's noise performance. Following this, the Master remained linked to the UTC(IT) reference, while the Slave sensors were connected to AHM signals provided by the FRATERNISE facility [5].

This configuration allowed for the evaluation of CTS's ability to estimate the relative frequency offset and drift of the AHM compared to UTC(IT). Based on these assessments, preliminary tests began to discipline the AHM versus UTC(IT) exploiting CTS measurements (see Fig. 1) and to disseminate a replica signal (UTC(IT)_CTS) remotely. This replica combines the short-term stability of the AHM with the medium- to long-term stability of UTC(IT).

Although still in its early stages, the results have been promising, demonstrating that CTS can produce a UTC(IT) replica with a stability of less than two nanoseconds within a 5-meter range. Subsequent studies are exploring the system's performance when one Slave is relocated 30 meters away, another is positioned 20 meters below ground, and a commercial Rubidium clock is used as a frequency signal reference for CTS Slaves.

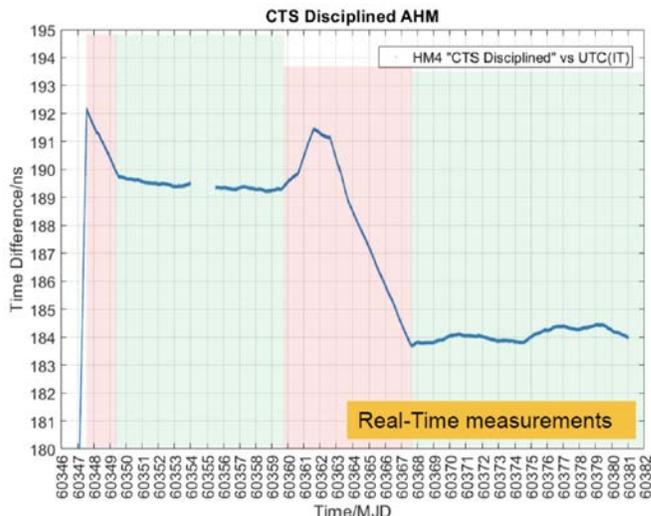

Fig. 1 First outcomes of disciplining atomic clocks using CTS. The plot shows the time difference between UTC(IT) and its remote reconstruction by disciplining an AHM through muon detection. Red areas indicate moments when the AHM system was not locked to UTC(IT) [6].

## V. CONCLUSIONS

Muochrony, an emerging scientific discipline, has shown significant promise but requires careful consideration of several factors before it can gain broad acceptance and standardization. From a metrological perspective, it is essential to thoroughly characterize the Cosmic Time Synchronizer (CTS) to evaluate its stability, accuracy, and robustness. Key steps include establishing a detailed uncertainty budget, developing specific calibration protocols, and improving the measurement technique and devices. These efforts are vital to position CTS as a traceable and reliable time dissemination method. CTS offers opportunities for standalone applications in critical environments or as a complementary solution to GNSS-based timing methods.